\documentclass[%
 reprint,
 amsmath,amssymb,
 aps,
prl,
superscriptaddress,
]{revtex4-2}

\DeclareUnicodeCharacter{2212}{\textminus}
\usepackage{graphicx}
\usepackage{dcolumn}
\usepackage{bm}
\usepackage[utf8]{inputenc}
\usepackage{amsmath}
\usepackage{xcolor}
\usepackage{microtype}
\usepackage{siunitx}
\DeclareSIUnit\atomicmassunit{u}
\DeclareSIUnit\bar{bar}
\usepackage{booktabs}
\usepackage{hyperref}
\usepackage{comment}
\usepackage{subcaption}

\begin{document}

\preprint{APS/123-QED}

\title{Demonstration of a mobile ion-trap quantum device}

\author{Sara Alfaro}
\affiliation{Universität Innsbruck, Institut für Experimentalphysik, Technikerstraße 25/4, 6020 Innsbruck, Austria}
\affiliation{Naturwissenschaftlich-Technische Fakultät, Universität Siegen, Walter-Flex-Straße 3, 57068 Siegen, Germany}

\author{Dipesh Bhanushali}
\email{dipesh.bhanushali@student.uni-siegen.de}
\affiliation{Naturwissenschaftlich-Technische Fakultät, Universität Siegen, Walter-Flex-Straße 3, 57068 Siegen, Germany}

\author{Sankarshan Choudapurkar}
\affiliation{Naturwissenschaftlich-Technische Fakultät, Universität Siegen, Walter-Flex-Straße 3, 57068 Siegen, Germany}

\author{Ashwin Bhartia}
\affiliation{Naturwissenschaftlich-Technische Fakultät, Universität Siegen, Walter-Flex-Straße 3, 57068 Siegen, Germany}

\author{Thomas Lafenthaler}
\affiliation{Universität Innsbruck, Institut für Experimentalphysik, Technikerstraße 25/4, 6020 Innsbruck, Austria}

\author{Tommaso Faorlin}
\affiliation{Universität Innsbruck, Institut für Experimentalphysik, Technikerstraße 25/4, 6020 Innsbruck, Austria}

\author{Yannick Weiser}
\affiliation{Universität Innsbruck, Institut für Experimentalphysik, Technikerstraße 25/4, 6020 Innsbruck, Austria}

\author{Lorenz Panzl}
\affiliation{Universität Innsbruck, Institut für Experimentalphysik, Technikerstraße 25/4, 6020 Innsbruck, Austria}

\author{Thomas Monz}
\affiliation{Universität Innsbruck, Institut für Experimentalphysik, Technikerstraße 25/4, 6020 Innsbruck, Austria}
\affiliation{Alpine Quantum Technologies, Technikerstraße 17, 6020 Innsbruck, Austria}

\author{Rainer Blatt}
\affiliation{Universität Innsbruck, Institut für Experimentalphysik, Technikerstraße 25/4, 6020 Innsbruck, Austria}
\affiliation{Alpine Quantum Technologies, Technikerstraße 17, 6020 Innsbruck, Austria}
\affiliation{Institute for Quantum Optics and Quantum Information of the Austrian Academy of Sciences, Technikerstraße 21a, 6020 Innsbruck, Austria}

\author{Giovanni Cerchiari}
\email{giovanni.cerchiari@uni-siegen.de}
\affiliation{Universität Innsbruck, Institut für Experimentalphysik, Technikerstraße 25/4, 6020 Innsbruck, Austria}
\affiliation{Naturwissenschaftlich-Technische Fakultät, Universität Siegen, Walter-Flex-Straße 3, 57068 Siegen, Germany}

\date{\today}

\begin{abstract}
Trapped-ion devices are being investigated for transportable applications, including the transfer of exotic charged particles between laboratories and the development of portable quantum technologies. To date, ion transport has been demonstrated only in magnetic (Penning) traps, whereas radiofrequency (Paul) traps are the predominant platform in quantum science. Here, we report the first transport of a few thousand ions confined in a linear Paul trap over distances of several kilometers, with transport durations of 15 and 45 min. These results establish a foundation for mobile trapped-ion quantum devices capable of operating while in motion and may enable the transport of exotic ions in Paul traps between experimental facilities.
\end{abstract}
 
\maketitle
 
 
\section{\label{sec:intro}Introduction}
 
Trapped ions constitute a leading platform for quantum information science, as quantum information can be encoded in their internal and motional degrees of freedom and manipulated with high precision~\cite{Zhu2022, Greganti2021}. Ion traps form the basis of trapped-ion quantum computers~\cite{Bruzewicz2019} and quantum-network nodes~\cite{Krutyanskiy2023} and provide a promising platform for quantum sensing~\cite{Cerchiari2021}. Despite achieving outstanding performance within compact footprints~\cite{Pogorelov2021}, trapped-ion quantum devices described as portable, including transportable optical clocks~\cite{Cao2017, Stuhler2021, Cao2022, Hao2024}, generally cannot operate while in motion. To date, only two notable demonstrations have overcome this limitation. More than three decades ago, a portable trap operating in motion was used to transport electrons~\cite{Tseng1993}. More recently, the transport of protons and antiprotons was demonstrated~\cite{Leonhardt2025, Leonhardt2026}. These experiments highlight the potential of mobile ion traps for transporting exotic particles, for example, to enable precision tests of matter-antimatter symmetry away from the noisy environment of accelerator facilities or to combine antimatter with exotic ions for nuclear studies~\cite{Aumann2022}.
 
Both demonstrations employed Penning traps, which confine charged particles using a combination of static electric and magnetic fields. Although Penning-trap-based quantum devices are being pursued~\cite{Jain2024, Pham2024, Hawaldar2024}, their use in quantum technologies remains limited compared with radiofrequency (RF) Paul traps, which constitute the predominant trapped-ion platform~\cite{Nop2021, Mazzanti2023,Hrmo2023,Delaney2024, OReilly2024, Hausser2025, Main2025, Mordini2025}. Interest in portable Paul-trap systems is growing, and such traps are already being investigated for future transportable implementations~\cite{FernandezGonzalvo2023, Chatterjee2025}. 
The requirements of this research differ markedly from those of portable ion-trap mass spectrometers~\cite{Yang2008, Hendricks2014}. Applications in quantum science and the transport of exotic ions require the trap to operate under ultrahigh-vacuum (UHV) conditions, retain ions for extended periods, and provide optical access for quantum operations. Here, we demonstrate the transport of ions confined in an apparatus designed to meet these requirements. The ions were transported over several kilometers while maintaining storage lifetimes comparable to those achieved under stationary laboratory conditions. This result opens opportunities for experiments involving laser cooling and quantum control in moving RF traps. Mobile trapped-ion systems could, for example, enable quantum-sensing experiments with accelerating clocks for tests of general relativity, as well as measurements of forces and magnetic fields. They could also facilitate the deployment of quantum-network nodes for distributed sensing. In addition, our apparatus incorporates an in-line all-metal angle valve that provides access for externally produced ions, offering a route toward loading and transporting exotic species by coupling the trap to an external ion beamline.
 
\section{\label{sec:setup} Experimental Setup}
 
The trap used in this study is a linear Paul trap in which radial confinement is provided by four parallel cylindrical electrodes of radius $r=\SI{1.75}{\milli\meter}$ and length $l=\SI{40}{\milli\meter}$. The distance $d$ between opposing electrodes is chosen such that $2r/d\approx1.145$, providing an optimal approximation to an ideal quadrupole potential~\cite{Pedregosa2010}. Two opposing electrodes are grounded, while the other pair is driven at an RF frequency $\Omega_{\mathrm{RF}}=2\pi\times\SI{17}{\mega\hertz}$ through a helical resonator. The resonator enhances the RF voltage amplitude by a factor of approximately 170. The voltage applied to the RF electrodes is monitored using a capacitive divider with an attenuation factor of $18.76(3)$.

\begin{widetext}
 
\begin{figure}[t]
    \centering
    \includegraphics[width=0.95\linewidth]{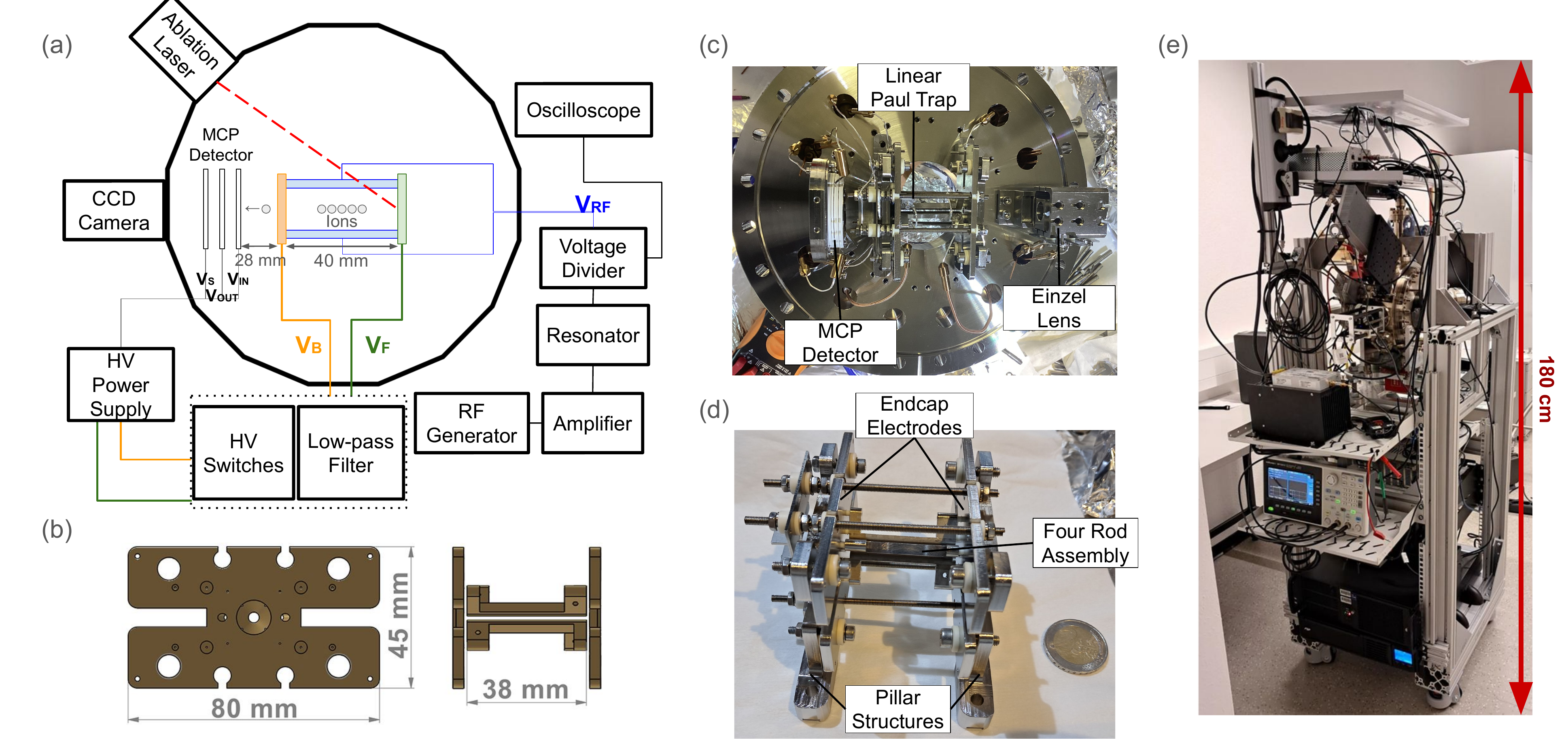}
    \caption{(a) Schematic of the experimental setup. Ions are produced by laser ablation, confined in a linear Paul trap, and detected using a microchannel plate (MCP) detector imaged by a camera. (b) Computer-aided design of the Paul-trap electrodes. The central aperture of diameter $\phi=\SI{3.4}{\milli\meter}$ in each planar endcap electrode serves as an exit port for extracting ions toward the MCP detector; the aperture is indicated by the red arrow. (c) Linear Paul-trap assembly mounted on a ConFlat-200 flange. An MCP detector is mounted along the trap axis for ion detection, while an Einzel lens is positioned on the opposite side to focus externally produced ions for injection into the trap. The Einzel lens was not used in the present study and is intended for future experiments involving ion injection from an external beamline. (d) Linear Paul-trap electrode assembly. Two opposing support structures secure the endcap electrodes and the four RF rods mounted between them. (e) Mobile Paul-trap rack housing the vacuum chamber and auxiliary equipment required to operate the trap during transport.}
    \label{fig:trap setup}
\end{figure}
 
\end{widetext}
 
Axial confinement is provided by two planar endcap electrodes, each featuring a central circular aperture of diameter $\phi=\SI{3.4}{\milli\meter}$. These apertures allow externally produced ions to be injected into the trap and trapped ions to be extracted for detection. A schematic of the trap and the overall experimental assembly is shown in Fig.~\ref{fig:trap setup}.
 
The trap is housed in a cylindrical vacuum chamber terminated by two ConFlat-200 flanges and equipped with 16 equally spaced ConFlat-40 radial ports. One of the ConFlat-200 flanges is custom designed to support the trap assembly and incorporates 12 additional ConFlat-16 ports for electrical feedthroughs. A microchannel plate (MCP) detector equipped with a phosphor screen is mounted along the trap axis on one side of the trap. The front surface of the MCP is oriented perpendicular to the trap axis and is located \SI{28}{\milli\meter} from the corresponding endcap electrode. Ion detection is performed destructively by extracting the trapped ions onto the MCP and recording the resulting phosphor-screen image with a digital camera (Hikrobot MV-CU013-A0GM).

The vacuum is maintained by an ion-getter pump (SAES Getters NEXTorr Z200) and a getter pump (SAES Getters CapaciTorr Z 400), mounted on two opposing ConFlat-40 ports of the chamber. A port aligned with the trap axis is equipped with a manual in-line all-metal valve, providing access to the vacuum system and enabling connection to an external ion source or beamline. The remaining ConFlat-40 ports are fitted with viewports to provide optical access to the trap.
 
The vacuum chamber is installed in a transportable rack that houses the equipment required for trap operation. This includes two high-voltage power supplies (iseg EBS EHS F020x and EHS 40150p) for the trap dc electrodes and MCP detector, respectively; an RF signal generator (Red Pitaya single-board computer); an RF amplifier (Mini-Circuits LZY-22+); and an oscilloscope (PicoScope 3405D MSO) for monitoring the RF voltage through the capacitive divider. Two uninterruptible power supplies (UPSs) provide power to the experimental apparatus during transport. The rack measures \SI{180}{\centi\meter} in height, \SI{61}{\centi\meter} in width, and \SI{90}{\centi\meter} in length. The trap and rack assembly contain no moving components, facilitating continuous operation during transport.
 
\section{\label{sec:methods}Experimental Procedure: Ion Generation \& Detection}
 
Ions are loaded into the trap by laser ablation using an ultra-compact Q-switched diode laser (Cobolt Tor XS, \SI{532}{\nano\meter}, \SI{50}{\micro\joule}). The ablation beam enters the vacuum chamber through a viewport and is focused to a spot smaller than \SI{100}{\micro\meter} on the endcap electrode opposite the MCP detector. The ablation position is optimized by generating ions and directing them toward the MCP. For this procedure, the RF field is applied while the axial confinement is removed, such that the trap acts as a quadrupole ion guide and directs positive ions toward the detector. The front face of the MCP is biased to approximately \SI{-1}{\kilo\volt} to enhance the collection of positive ions.
 
The ion yield from laser ablation varies over time, requiring the ablation spot to be repositioned approximately every five to six loading attempts to maintain a detectable signal at the MCP. We attribute this variation to a combination of laser beam-pointing instability and modification of the electrode surface induced by repeated ablation. Once the ablation position is optimized, ions are loaded by applying ten laser pulses while both endcap electrodes are biased to positive potentials, providing axial confinement for positively charged ions produced by ablation. Ten pulses were chosen as a compromise between reducing shot-to-shot fluctuations in the ablation yield and maintaining a stable vacuum pressure below $\SI{3e-11}{\milli\bar}$, which minimizes fluctuations in the MCP background count rate.
 
For the measurements reported below, the endcap potential is set to \SI{1}{\kilo\volt}, while the RF electrodes are driven at an amplitude of \SI{600}{\volt} and a frequency of \SI{17.1}{\mega\hertz}. The measurements used to determine these operating parameters are presented in the Supplemental Material. Under these conditions, we trap ions with a mass-to-charge ratio of approximately \SI{12}{\atomicmassunit}$/e$, which we attribute predominantly to singly charged carbon ions originating from the stainless-steel electrode. In a separate experiment, we verified confinement of ion species with mass-to-charge ratios of approximately \SI{50}{\atomicmassunit}$/e$ by operating the trap at \SI{12.04}{\mega\hertz}. These results indicate that the platform can accommodate a broad range of ion species, including species commonly employed in trapped-ion quantum-information experiments, such as Be$^+$, Ca$^+$, and Sr$^+$.
 
\begin{figure}[t]
    \centering
    \includegraphics[width=\linewidth]{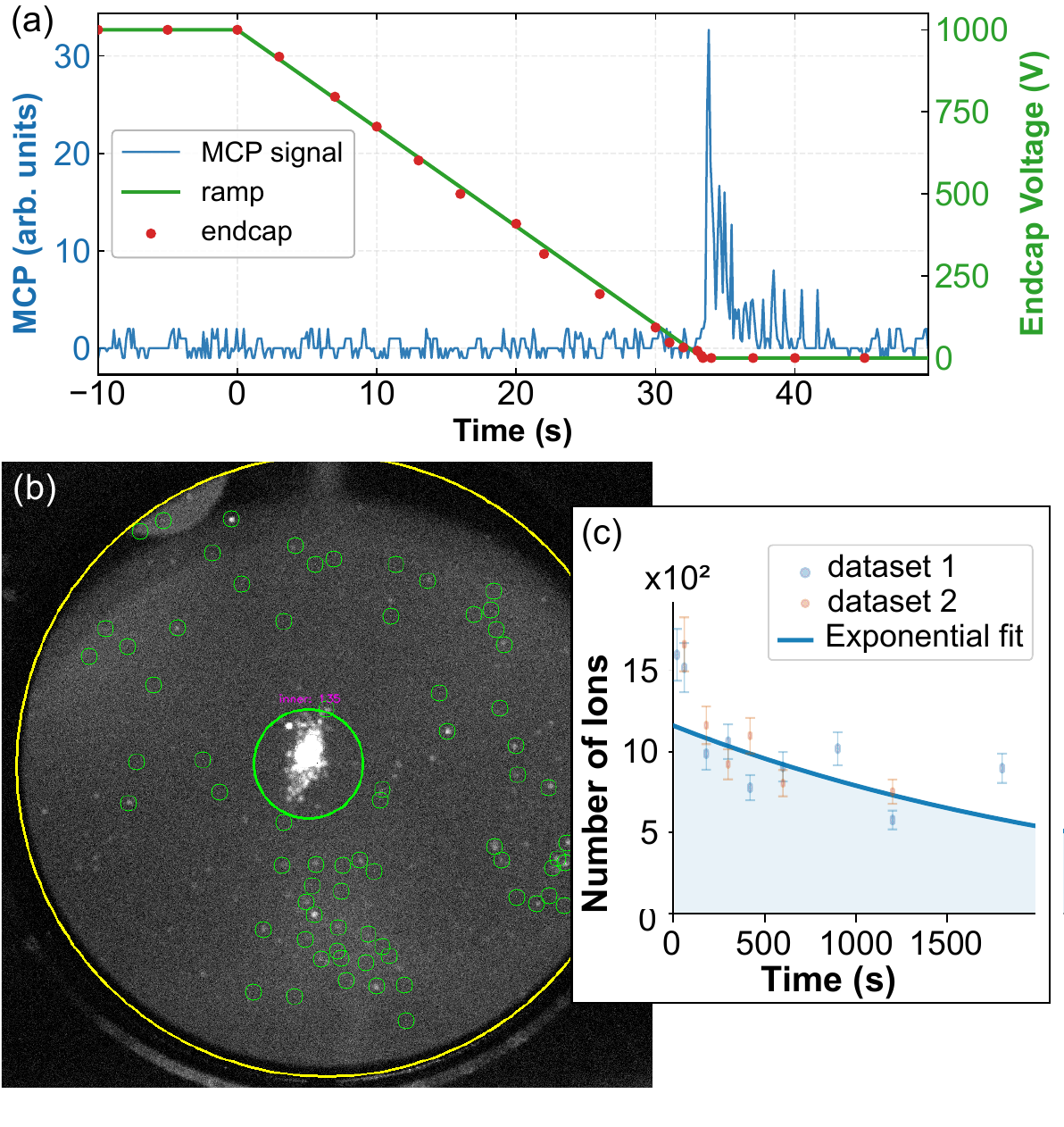}
    \caption{\textbf{Examples of experimental data and ion detection.} (a) Integrated MCP signal (blue line, left axis) and corresponding endcap voltage (red points, right axis) as a function of time during ion extraction from the trap. A detectable ion signal appears only when the endcap voltage decreases below approximately \SI{10}{\volt}. (b) Representative MCP image illustrating a worst-case condition for the image-analysis procedure, with strong background illumination. The image is divided into a central region (inside the green circle), where ion signals exhibit pileup, and an outer annular region (between the green and yellow circles), where individual ion events remain spatially resolved (small green circles). The outer yellow circle has a diameter of \SI{25}{\milli\meter}. Ions in the single-ion region are counted individually, and their integrated signals are used to estimate the number of ions in the central pileup region, as described in the main text. (c) Number of trapped ions as a function of storage time measured under stationary laboratory conditions. An exponential fit yields a storage lifetime of $\tau = 2.6(5)\times10^3$~\si{\second}.}
    \label{fig:ionrelease}
\end{figure}
 
The endcap electrodes are connected to the power supply through low-pass filters with a time constant of a few milliseconds to suppress electrical noise that could perturb the trapped ions. Ions are extracted toward the MCP by setting the voltage of the endcap facing the detector to \SI{0}{\volt} using the power-supply controller. An example of the ion-extraction sequence is shown in Fig.~\ref{fig:ionrelease}(a), where the integrated MCP signal and the voltage applied to the endcap electrode are plotted as a function of time. Ion extraction occurs only after the endcap voltage decreases below approximately \SI{10}{\volt}, consistent with a radial trap depth of a few electron volts.

The integrated MCP signal is obtained from videos recorded with a resolution of $1280\times1024$ pixels and an 8-bit intensity depth. A representative frame is shown in Fig.~\ref{fig:ionrelease}(b) to illustrate the image-analysis procedure. Individual MCP images are extracted from the recorded video and preprocessed to remove systematic background contributions. A reference background image is constructed by randomly sampling 300 frames from the video and calculating the median intensity at each pixel, $I_{\mathrm{bg}}(x, y) = \operatorname{median}\left\{ I_i(x, y) \right\}_{i=1}^{300}$. This median-based procedure effectively suppresses stochastic background fluctuations as well as transient ion signals, which occur at varying positions across the detector. Background-subtracted images are obtained according to $I_{\mathrm{diff}}(x,y) = \max\left[0,\; I_{\mathrm{frame}}(x,y) - I_{\mathrm{bg}}(x,y)\right]$.

For measurements exhibiting signal pileup near the center of the MCP, each image is divided into two regions: a central pileup region, where individual ion events overlap, and a single-ion region, where individual events remain spatially resolved. In the single-ion region, events are identified using a local-maximum detection algorithm subject to three criteria: a peak intensity exceeding $12$~a.u.\ in a single pixel, an integrated intensity exceeding $150$~a.u.\ within a $7\times7$-pixel window, and a minimum separation of 15 pixels between detected events. This automated procedure provides a reproducible method for identifying and counting spatially resolved single-ion events.

The number of ions in the pileup region is estimated by dividing the integrated signal in this region by a calibrated single-ion signal. To assess the associated systematic uncertainty, the single-ion signal is determined using two independent methods. In the first (average) method, the mean integrated intensity within an $11\times11$-pixel window centered on each detected single-ion event is used to determine the average signal per ion. In the second (median) method, the single-ion signal is obtained from the median of the integrated intensities within $7\times7$-pixel windows centered on individual events. The two methods yield identical ion-count estimates only for an appropriate symmetric distribution of single-ion signals. In practice, the average method yields systematically higher ion counts, with a mean relative difference of +9\%, while the root-mean-square (rms) difference between the two estimates for the central region is 73.6\%. For the 15-min transport measurement, this difference corresponds to a relative uncertainty of 33\% in the total ion count. We therefore use the difference between the two calibration methods to estimate the systematic uncertainty in the ion number for measurements exhibiting pileup, accounting for the nonlinear response of the detector at high local ion densities.
 
The ion storage lifetime measured under stationary laboratory conditions is shown in Fig.~\ref{fig:ionrelease}(c). The measured lifetime, of order $10^3$~s, is consistent with values reported for similar traps~\cite{Cerchiari2019}. A relative uncertainty of 10\% is assigned to the measured ion number based on repeated measurements performed at the same storage time. We attribute this variation primarily to shot-to-shot fluctuations in the ion-loading process.
 
For the ion-transport measurements, the rack containing the trap and supporting equipment was loaded into a van using a forklift. The rack was secured to the floor of the vehicle with ratchet straps, and its wheels were immobilized using the integrated support feet. During transport, the entire experimental apparatus was powered exclusively by the UPSs. Each transport measurement followed the same procedure. First, with the van stationary, ions were generated by laser ablation and confined in the trap at the designated ``Load'' location. The van was then driven over a distance of several kilometers while the ions remained trapped. Upon reaching the designated ``Measure'' location, the van was stopped and parked to provide access to the experimental apparatus. Finally, the trapped ions were extracted toward the MCP detector by setting the voltage of the endcap facing the detector to \SI{0}{\volt}.

\begin{figure}[t]
    \centering
    \includegraphics[width=\linewidth]{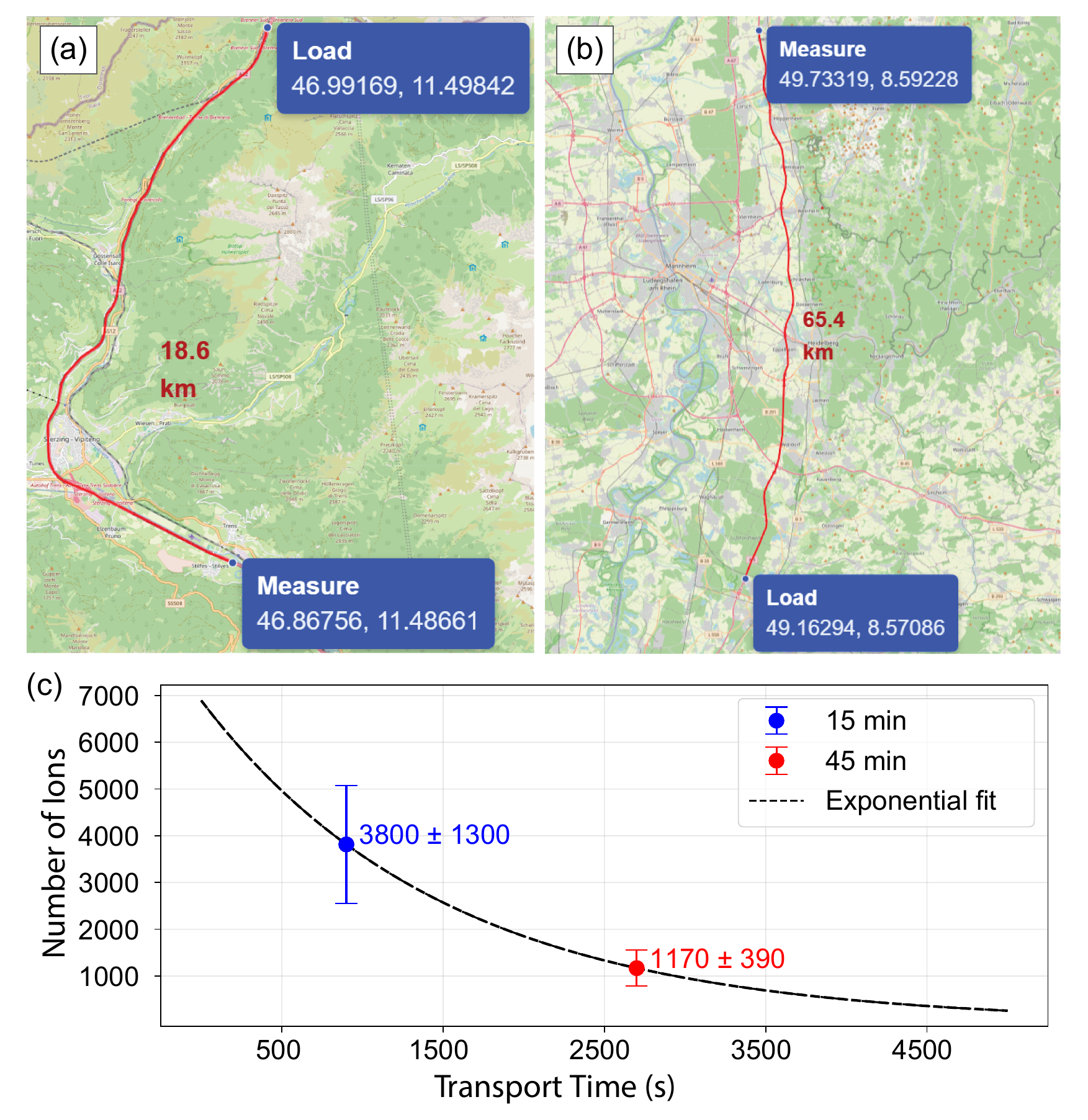}
    \caption{\textbf{Ion-transport measurements.} Transport routes corresponding to storage times of (a) \SI{15}{\minute} and (b) \SI{45}{\minute}. North is oriented upward in both maps. (c) Number of ions detected after transport as a function of storage time. An exponential fit yields a storage lifetime of $\tau = 1.52(71)\times10^3$~\si{\second}.}
    \label{fig:transportmeasure}
\end{figure}

The mobile-operation measurements are summarized in Fig.~\ref{fig:transportmeasure}. The transport routes were chosen along highways and included both straight and curved sections to provide representative driving conditions. Throughout each measurement, from ion loading to detection, the camera monitoring the MCP detector recorded continuously. This allowed the storage time to be determined from the time interval between the ablation signal and the subsequent detection of the stored ion cloud. In a first transport measurement, ions were loaded at the Lanz-Brenner rest area, shortly after crossing the Austria--Italy border (approximate coordinates: 46.9958, 11.5045). The van containing the operating trap was then driven for \SI{18.6}{\kilo\meter} before stopping at Campo di Trens (approximate coordinates: 46.8807, 11.4812). After a total storage time of \SI{15}{\minute}, the ions were extracted onto the MCP detector. We measured $3.82(6)(1.3)\times10^3$ ions, where the first uncertainty is statistical and the second is systematic. In the second measurement, ions were loaded near Bruchsal (approximate coordinates: 49.1611, 8.5929). The van was subsequently driven for \SI{63.4}{\kilo\meter} before stopping near Zwingenberg (approximate coordinates: 49.7331, 8.5922), where the ions were extracted after a storage time of \SI{45}{\minute}. We measured $1.17(3)(3.9)\times10^3$ ions, corresponding to an approximately 70\% reduction relative to the 15-min measurement. This decrease is consistent with the ion storage lifetime measured under stationary laboratory conditions. The reported coordinates are specified to a precision of approximately \SI{11}{\meter}.
 
\section{\label{sec:conclusion}Conclusion}
We have demonstrated the transport of a few thousand positively charged ions confined in a radio-frequency linear Paul trap over distances of several kilometers. The ions remained trapped for 15 and 45~min while the apparatus was transported in a moving vehicle, with storage lifetimes consistent with those measured under stationary laboratory conditions.
 
Our results establish a complementary route toward mobile trapped-ion quantum technologies by demonstrating ion confinement during motion. A natural next step toward a fully operational mobile quantum device is to demonstrate laser cooling and coherent control of trapped ions while the apparatus is in motion. Such capabilities could enable mobile quantum sensors and transportable quantum-network nodes operating during transit.
 
Our work also complements ongoing efforts to transport exotic charged particles, including antimatter, by establishing a pathway toward their transport in RF traps. This approach could ultimately enable precision measurements of exotic particles outside the strong magnetic fields required for confinement in Penning traps. Operation in a low-magnetic-field environment could improve experimental accessibility and reduce magnetic-field-induced energy shifts and splittings that can complicate high-precision spectroscopy.

\begin{acknowledgments}
\textit{Acknowledgments}. This work was supported by the Ministry of Culture and Science of North Rhine-Westphalia. The authors would like to thank Atharva Bahekar for helping characterizing the helical resonator, Peter Tirler for sharing with us the measurement on the German highway, and Ulrich Werthenbach for designing the support of the camera looking at the MCP. The authors are thankful to Nik Röcher, Olaf Meyer and Jörn Falke for the mechanical manufacturing of the trap's electrodes and the support of the vacuum chamber. The authors would like to thank Roberto Brusa for hosting the trap over at the University of Trento a few days during the presented measurement campaign.
\end{acknowledgments}

\bibliography{refs}

\appendix
\newpage
\section{Supplementary material}

\begin{figure}[h]
    \centering
    \includegraphics[width=\linewidth]{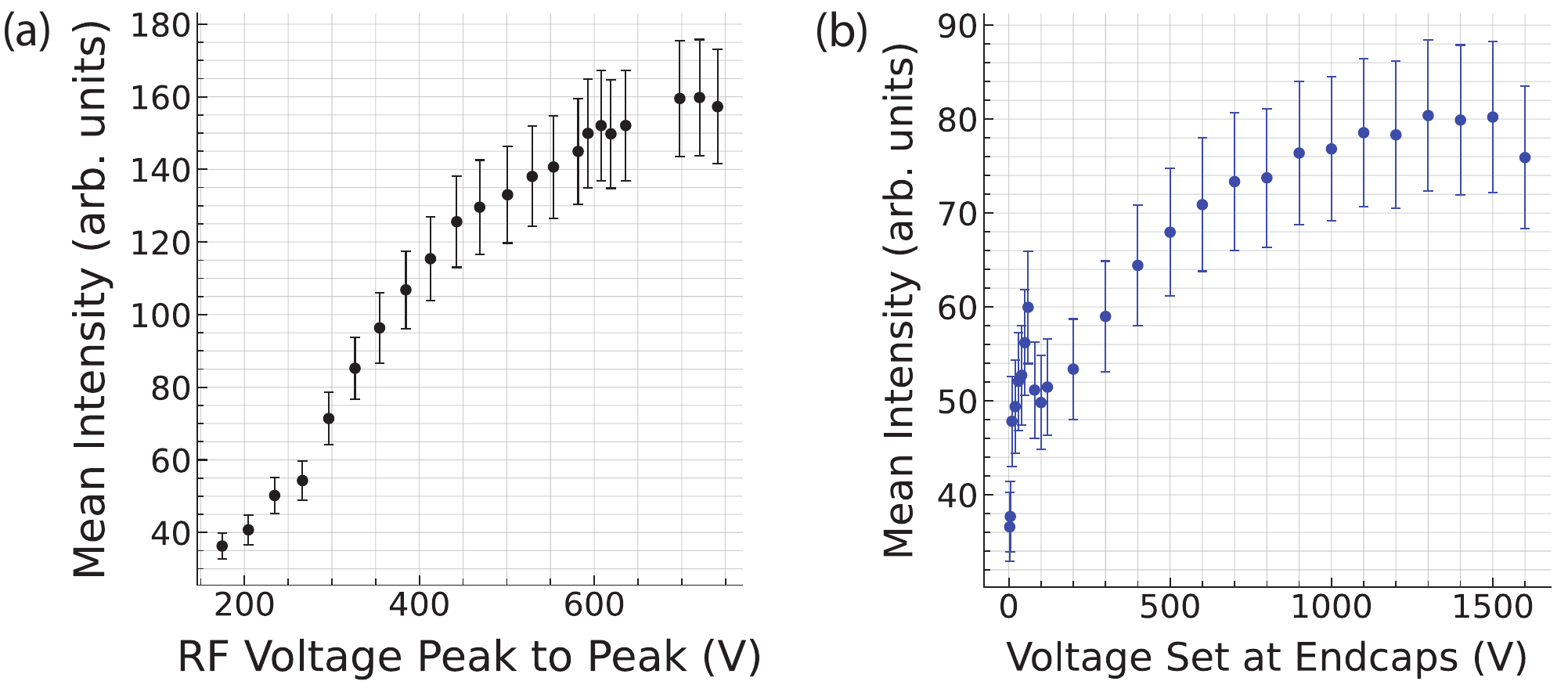}
    \caption{(a) Detector signal as a function of the RF peak-to-peak amplitude at constant voltage applied to the endcaps. (b) Detector signal as a function of the voltage applied to the endcaps at constant RF drive. The scans shows the range accessible by the experiment.}
    \label{fig:Figsupp}
\end{figure}

The voltage of the RF field and the voltage value of the endcaps were optimized via the laboratory measurements presented in Fig.~\ref{fig:Figsupp}. For these optimizations, we connected the endcaps via two high-voltage switches (Behlke-HTS 61-01-GSM), which allowed us to quickly release the trapped ions onto the detector after \SI{30}{\second} storage time. For the experiments reported in the main text, the endcap voltage of \SI{1000}{\volt} and a RF peak-to-peak amplitude of \SI{600}{\volt} were used.

\end{document}